# Tantalum STJ for Photon Counting Detectors

Corentin Jorel, Philippe Feautrier, Jean-Claude Villégier

*Abstract*— Superconducting Tunnel Junctions (STJ's) are currently being developed as photon detectors for a wide range of applications. Interest comes from their ability to cumulate photon counting with chromaticity (i.e. energy resolution) from the near infrared (2 µm) to the X-rays wavelengths and good quantum efficiency up to 80%. Resolving power can exceed 10 in the visible wavelength range. Our main goal is to use STJ's for astronomical observations at low light level in the near infrared.

This paper put the emphasis on two main points: the improvement of the tantalum absorber epitaxy and the development of a new version of the fabrication process for making Ta/Al-AlOx-Al/Ta photon counting STJ's. The main features of this process are that pixels have aligned electrodes and vias patterned through a protecting $SiO_2$ layer. These vias are then used to contact the top electrode layer. We use a double thin aluminum trapping layer on top of a 150 nm thick Ta absorber grown epitaxially. Photon counting experiments with Ta junction array are presented at $\lambda$= 0.78 µm. Digital filtering methods are used to compute the photon counting data in order to minimize the effects of noise.

## I. INTRODUCTION

Superconducting Tunnel Junctions (STJ's) have already demonstrated their capability to count photons in the near infrared with a moderate energy resolution [1]-[2]. Astronomers now use STJ's in photon counting cameras that have an intrinsic and moderate energy resolution [3].

The aim is to investigate the interest of this kind of detectors for ground based astronomical applications and compare them to Hot Electron Bolometers devices (or Superconducting Single Photon Detectors, SSPD) that are also developed in our group [4]. Since the conventional detector arrays based on semi-conductor devices have recently progressed toward the direction of photon counting detectors in the visible [5], some niches have to be found for this type of detectors where conventional devices cannot compete. The astronomical applications that could be investigated are wave-front sensing detectors for adaptive optic systems and interferometry (fringes sensors, focal plane instruments) in the near infrared.

## II. JUNCTIONS FABRICATION

We have developed an original fabrication process of Ta (150 nm) /Al-AlOx(1nm)-Al (80 nm)/ Ta (120 nm) superconducting junctions [6]. The multilayer is deposited in-situ by DC-Magnetron sputtering. The aluminum oxide barrier is obtained by static oxidation (200 mbar of pure oxygen atmosphere during 30 minutes). R-plane polished Sapphire is used as substrate since the square section presented on surface is particularly adapted to the cubic lattice of Ta. Since the Al energy gap is smaller than the Ta one (0.172 and 0.664 meV), the two aluminum layers are used to support the localization of the excited quasi-particle charges near the oxide barrier.

### A. Absorber epitaxy

Like a majority of photon counting detectors, STJ converts the incident energy into an excited charge population which gives an access to the deposited energy. The photon-excited charges tunnel through the oxide barrier and the energy can be measured by integration of the photo-current. Thus, the crystalline quality of the Ta base-electrode (called "absorber"), in which the photons are absorbed, is a critical parameter and we need to maximize the lifetime of the photo-excited quasi-particles.

Standard Ta dc-magnetron sputtering on Nb buffer layer, in a background pressure of about $10^{-7}$ mbar and at room temperature, gives a polycrystalline film as shown on upper θ-2θ, X-rays diffraction scan (see Fig. 1). By sputtering the film directly on the R-plane sapphire heated at 600 °C, we obtained a much more textured film but the orientation inside the growth plan was still too dispersed and a diffraction peak intensity is not highly improved. Therefore we used a thin Nb buffer layer (<10 nm) and a high substrate temperature to grow epitaxially the right cubic Ta phase (this Nb buffer layer is expected to facilitates the nucleation of the right cubic Ta lattice on R-plane sapphire). We obtained a very significant improvement of the diffraction peak intensity. Using Scherer's law, we estimated diffracting domain sizes larger than 90 nm and Ta (200) rocking-curve XRD gives 0.3° of grain disorientation angle with the growth axis.

The RRR measurement (Relative Resistive Ratio) confirms the improvement of the Ta quality, the typical mean free path

Manuscript received October 5, 2004. This work was funded by the CNES French Space Agency under a R&T program in Astronomy.

Corentin Jorel is with the Laboratoire d'Astrophysique de l'Observatoire de Grenoble UJF-BP 53X-38041 Grenoble Cedex, France, and with the CEA-Grenoble SPSMS/LCP, F38054 Grenoble Cedex 9, France (e-mail: jcorentin@yahoo.fr).

Philippe Feautrier is with the Laboratoire d'AstrOphysique de Grenoble UJF-BP 53X-38041 Grenoble Cedex, France (corresponding author, e-mail: Philippe.Feautrier@obs.ujf-grenoble.fr phone: +33-476635981; fax: +33-476448821;).

Jean-Claude Villégier is with CEA-Grenoble SPSMS/LCP, F38054 Grenoble Cedex 9, France (e-mail: villegier@cea.fr).

under 10K $l_{10K}$ was increased by a factor 10: $l_{10K} = 160$ nm.

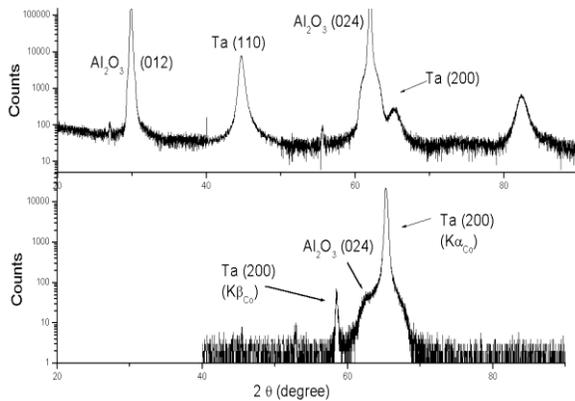

Fig. 1. θ-2θ XRD scan of 200 nm thick sputtered Ta thin film on R-plane sapphire substrates covered in-situ by a very thin (<10 nm) Nb buffer layer. Up: polycrystalline Ta film sputtered at room temperature. Down: epitaxial Ta film deposited on substrate heated at 600°C.

### B. The fabrication process steps

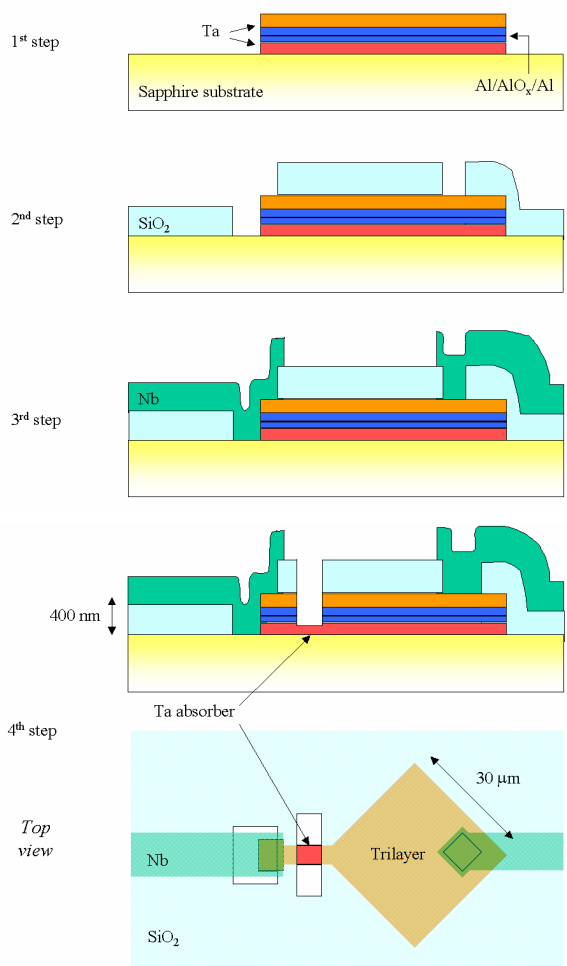

Fig. 2. The different fabrication process steps (side view) and top view of the final Ta-STJ device.

The main four steps of the process are illustrated on Fig. 2 and involve classical microelectronics procedures. The junctions and absorber area are defined by the first photolithography mask. Tantalum layers are selectively etched by Reactive Ion Etching (RIE) in a $SF_6$-$O_2$ gas mixture.

The etch of the aluminum tri-layer is critical since the barrier is very thin (of the order of one nm) and may be easily shunted at the junction edge by micro-bridges of etching residues. Aluminum RIE is very difficult with fluorinated gases and therefore requires the use of chlorinated gas that can not be presently handled in our clean room for safety reasons. Ion milling was used in our previous niobium-based generation of STJ [7], but with this etching technique, the obtained proportion of good-quality Ta junctions was found very poor. That is the reason why we decided to use a wet isotropic etching solution of phosphoric acid, nitric acid, acetic acid and water (respectively 85%-25 Volumes, 70%-1Vol., 99.7%-5Vol., 2Vol.). This solution removes the Al film within 15 to 20 min. with an excellent material selectivity and without any observed mechanical film damages. A microscopic inspection was used to confirm the visual determination of the etch end. The inherent drawback of this technique is the etching isotropy: aluminum lateral over-etch can lead to leakage current coming from the junction's edges. The second mask allows to produce an electrical contact to the Ta absorber by the etching of the $Ta_{sup}$ / Al-AlO$_x$-Al layers on a narrow strip with the same etching techniques. Immediately after this step, an MgO/SiO$_2$ (6 nm/400-500 nm) insulating film is sputtered on a pre-patterned photoresist having a negative profile for an easy lift-off of the contact holes. The last two masks allow the definition of the Nb current leads by RIE and Ti-Au contact pads by lift-off.

The main motivation for developing this new version of the process was the self-alignment of the electrodes during the first etching, which prevents the absorber from having air-exposed surfaces on which many recombination sites appear. In that way we reduce the recombination loss of the photo-generated charges. Moreover, the use of a SiO$_2$ insulating layer immediately after the junction definition protects their edges from the ambient air damages and all further process steps.

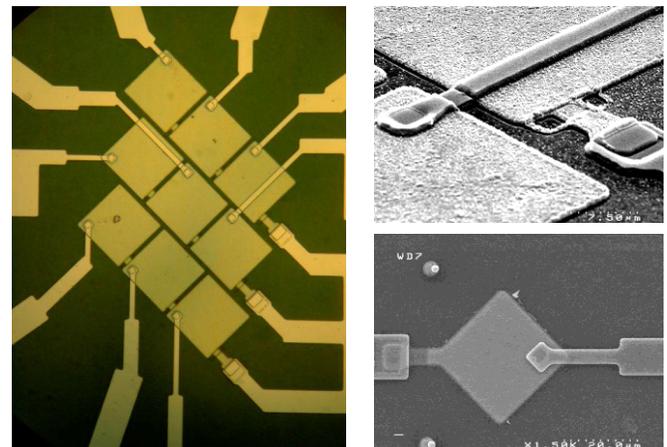

Fig 3. Left: 3x3 pixel array of 30 μm x 30 μm Ta-based STJ's; upper right: details of the array structure with a SEM; lower right: single Ta-STJ

The Fig. 3 shows some pictures of produced STJ's with this original process. 3x3 pixel arrays have been produced as shown on this figure (left). Also shown on this figure are pictures of a single detector and details of the junction structure (see Fig. 3, bottom right hand side).

### III. JUNCTION CHARACTERISATION

Several wafers were fabricated using this original junction fabrication process. This process is now reliable and not strongly sensitive to junction fabrication drifts. For the first time, we obtained reproducible junctions without a large dispersion of the characteristics (I/V characteristics, junction quality, leakage current, normal resistance…).

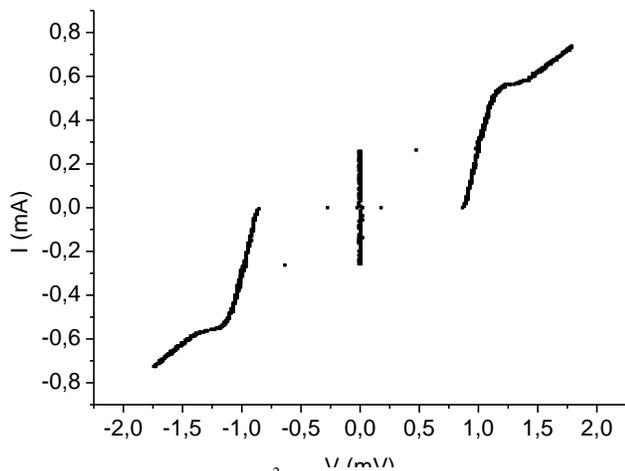

Fig 4: I-V curve of a 40x40 µm$^2$ Ta STJ from a 3x3 pixel array, the normal resistance is 2.4 Ω, the current density is 16.25 A/cm$^2$. The temperature is 50 mK.

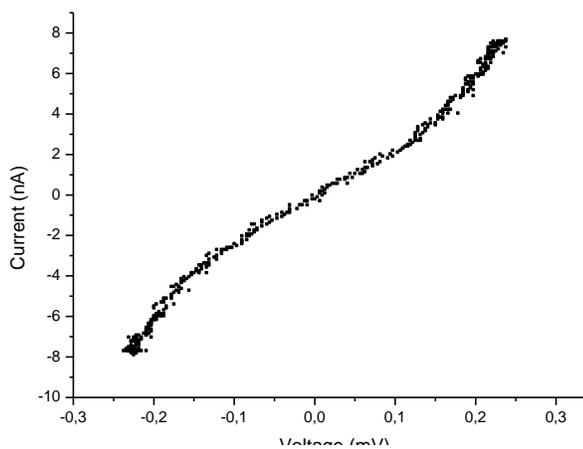

Fig. 5: Sub-gap leakage current of the Ta STJ from the Fig.4. The dynamic resistance is 50 kΩ at 0.1 mV. The temperature is 100 mK.

Fig. 4 shows a typical I-V curve of tantalum junctions from a 3x3 pixel array, measured at a temperature of 50 mK. For a junction area A of 40x40 µm$^2$, the normal resistance $R_n$ is 2.4Ω and the product $R_n.A$ is 38.4 µΩ.cm$^{-2}$, which is not transparent enough to expect good photon counting ability. This is the reason why the next wafers to be tested as photon counting detectors are oxidized in reduced oxygen pressure, 1 mbar during 50 min, compared to the device presented here oxidized at 200 mbar during 30 min. Fig. 5 shows the corresponding I-V curves under the gap voltage from which is derived the dynamical resistance (Rd, i.e. leakage current) at the bias voltages used during photon detection, typically 0.1mV for our devices. The $R_d$ value at 0.1 mV is 50 kΩ here. The electrical uniformity in an 3x3 pixels array has been studied and Fig. 6 presents an I-V curves superposition with a 15 % normal resistance dispersion.

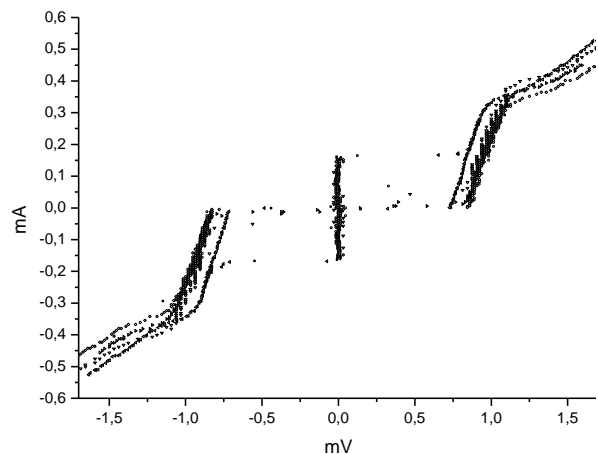

Fig. 6: I-V curves superposition of the different junctions of a 3x3 pixel array. Normal resistances of the junctions are close to 30 µΩ/cm$^2$ at 150 mK. (mean value ~ 31 µΩ/cm$^2$ and standard deviation of 5 µΩ/cm$^2$). Corresponding dynamic resistances are in the 0.15 -0.5 Ω.cm$^2$ range.

### IV. SINGLE PHOTON DETECTION

The same Ta-based junction has been used for optical photon counting at a temperature of 0.15K. The light source is a pulsed photodiode emitting in the near infrared (λ=0.78 µm). This is an ABB Hafo 1A330 device with an output power of 100 µW and a rise/fall time of 15 ns. The photodiode is placed inside the cryostat and is pulsed with an electrical signal of a few microseconds sent from the outside. The photodiode is coupled to a single mode optical fiber having a 5 µm diameter core through a commercial SMA connector. This fiber is then fixed to the cold finger where the junction is placed, illuminating the back-side of the junction. Since the optical fiber tip is placed one centimeter away from the junction and the optical fiber diameter is very small, the coupling efficiency of the photodiode light to the junction is intentionally low in order to reduce the photon flux. This experimental set-up allows us to illuminate the junction with very few photons at each photodiode pulse by dilution of the optical fiber output beam. In addition, classical digital filtering methods are used to compute the photon counting data in order to minimize the effects of noise.

The spectrum recorded shows several peaks having a charge proportional to the number of photons absorbed depending on the statistical fluctuations of the number of photons emitted (see Fig. 7). The interval between two consecutive peaks is uniform and corresponds to the mean charge <Q(λ)> produced by the absorption of one single photon. This demonstrates the photon counting ability of our experimental set-up in the near infrared, even if the signal versus noise ratio is not optimal and could be improved by decreasing the junction tunneling time (i.e. normal resistance in our case). In this figure, the mean number of photons illuminating the junction is 6, but the photon peaks with 5, 7 and 8 photons can also be detected.

From this figure, the mean charge $<Q(\lambda)>$ created by one single photon absorption is estimated by the mean difference between two consecutive peaks :

$$\langle Q(\lambda) \rangle \approx 2800 \; e^-.$$

If we consider that for a Ta absorber, the initial number of quasi-particles $N_0$ is given by :

$N_0 \approx 7.1 \times 10^5 / \lambda \, \Delta(T)$, where:

$\lambda$ = photodiode wavelength (nm) = 780 nm,

$\Delta(T)$ = energy gap of the Ta-Al electrode (meV) = 0.45 meV,

we can then calculate $N_0$ for our junctions : $N_0 \approx 2000 \; e^-$.

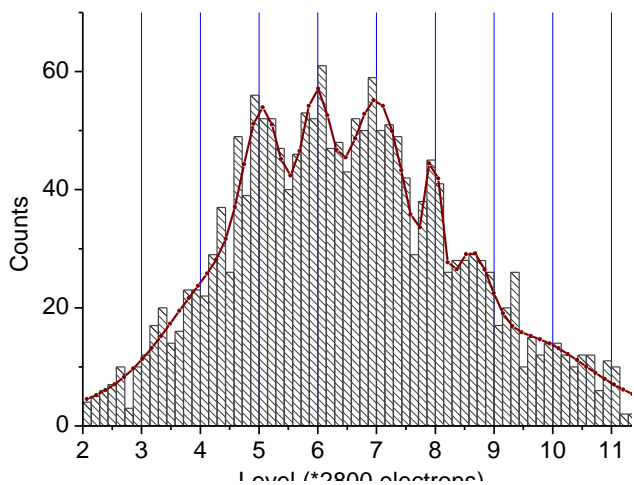

Fig. 7. Photon counting spectrum at $\lambda=0.78$ μm with a Ta-based STJ. The horizontal axis was normalized to the mean number of collected charge created by one photon.

This allows us to estimate how much time each quasi-particle originally created in the absorber film is transferred across the barrier, in the absence of loss processes:

$$\langle n \rangle \approx \langle Q(\lambda) \rangle / N_0 \approx 1.4$$

This means that with our junctions and their corresponding parameters (low transparency), no real multiple tunneling phenomenon of the quasi-particles through the tunnel barrier is observed.

It is also possible to derive the noise of the amplifier chain coupled to this junction from the numerical pulse data. RMS noise of 700 $e^-$ has been computed: it is the quarter of the charge $<Q(\lambda)>$ measured for one single photon detected at $\lambda=0.78$ μm.

## V. CONCLUSION

This paper describes an original Ta STJ fabrication process optimized for our particular application of optical photon counting detectors. This new fabrication process gives very reliable devices and hardly depends on fabrication parameters. This process allows the fabrication of Ta STJ arrays with a Ta absorber having good mean free path and a large quasi-particle lifetime, which are both required for a good detector performance. The Ta absorber quality was fully studied. By adding an original Nb buffer layer, we demonstrated that the absorber quality can be easily improved and is less sensitive to the sputtering conditions: background pressure and sputtering temperature.

The use of digital techniques also allowed to minimize the noise effects from our data. Current results are promising and main improvements should come from an increase of the barrier transparency (i.e. by decreasing the barrier thickness) which increases also the quasi-particle lifetime. This is expect to be achieved without significant degradation of the junction quality, and in particular without increasing the leakage current of the STJ.

We are currently investigating the use of this type of detector for ground-based astronomy in the near infrared. Instruments requiring the lowest achievable noise, as well as a good quantum efficiency and high rates at the same time should benefit from this type of detectors. Among them, Adaptive Optics wave-front sensors could be a good candidate to test this new concept.

## ACKNOWLEDGMENT

The authors would like to thank J-L. Thomassin, I. Schuster from CEA-G/DRFMC, for their contributions to the fabrication and characterizations of the devices and A. Benoît, S. Leclercq, J-L. Bret and P. Camus from CNRS-CRTBT, for their support to the detection experiments.